\documentclass[superscriptaddress,prl,amsmath,twocolumn,amssymb]{revtex4}

\usepackage{dcolumn}
\usepackage{amsfonts}
\usepackage{amssymb}
\usepackage{amsmath}
\usepackage{amsthm}
\usepackage{latexsym}
\usepackage{epsfig}
\usepackage{color}
\usepackage{lscape}
\usepackage{multirow}
\usepackage{braket}
\usepackage{bbm}
\usepackage{mathrsfs}
\usepackage{fancyhdr}
\usepackage{lastpage}
\usepackage{tikz}
\usetikzlibrary{backgrounds,decorations.pathreplacing,calc,arrows,shapes,fit}
\input{Qcircuit.tex}

\newcommand{\be}{\begin{equation}}
\newcommand{\ee}{\end{equation}}
\newcommand{\ba}{\begin{array}}
\newcommand{\ea}{\end{array}}
\newcommand{\bea}{\begin{eqnarray}}
\newcommand{\eea}{\end{eqnarray}}

\newcommand{\mb}[1]{\boldsymbol{\mathbf{#1}}}
\newcommand{\llg}{\langle\langle}
\newcommand{\rrg}{\rangle\rangle}

\begin{document}

\title{Derandomizing quantum circuits with measurement based unitary designs}

\author{Peter S. Turner}
\email{peter.turner@bristol.ac.uk}
\affiliation{School of Physics, H. H. Wills Physics Laboratory, Tyndall Avenue, University of Bristol, Bristol BS8 1TL, UK.}
\date{\today}

\author{Damian Markham}
\email{markham@enst.fr}
\affiliation{CNRS LTCI, Departement Informatique et Reseaux, Telecom ParisTech, 23 avenue d'Italie, CS 51327, 75214 Paris CEDEX 13, France}

\date{\today}

\begin{abstract}
Entangled multipartite states are resources for universal quantum computation, but they can also give rise to ensembles of unitary transformations, a topic usually studied in the context of random quantum circuits.
Using several graph state techniques, we show that these resources can `derandomize' circuit results by sampling the same kinds of ensembles quantum mechanically, (analogously to a quantum random number generator).
Furthermore, we find simple examples that give rise to new ensembles whose statistical moments exactly match those of the uniformly random distribution over all unitaries up to order $t$, while foregoing adaptive feed-forward entirely.
Such ensembles -- known as $t$-designs -- often cannot be distinguished from the `truly' random ensemble, and so they find use in many applications that require this implied notion of pseudorandomness.  
\end{abstract}

\maketitle

\emph{\textbf{Introduction --}}
Randomness is an important resource in both classical and quantum information theory, underpinning cryptography, characterisation, and simulation.
Random unitary transformations are often considered in the form of random quantum circuits, with wide-ranging applications in, for example, estimating noise\cite{benchmarking}, private channels\cite{Hayden..Winter}, modelling thermalisation\cite{Muller..Wiebe}, photonics\cite{Matthews..Turner}, and even black hole physics\cite{HaydenPreskill}.
Uniform randomness, sampling from the `flat' Haar measure on a continuous group, is however very resource intensive.
A natural definition of a less costly \emph{pseudo}random ensemble is one whose statistical moments are equal to those of the Haar ensemble up to some finite order $t$ -- this is the defining property of a $t$-design.
Arising in classical coding theory\cite{DelsarteGoethalsSeidel}, in the quantum community designs were first applied to states\cite{RenesBlumekohoutScottCaves}, and later to processes\cite{Dankert}, the latter being our concern here.
As mathematical objects they are of interest in their own right, not least of which as generalisations of SICPOVMs and MUBs which provide infamous open problems\cite{Appleby-Durt}. 

Rather than quantum circuits composed of sequences of gates, unitary transformations in a measurement based (MB) model\cite{RaussendorffBriegel} are realized by sequences of measurements on highly entangled resource states.
These have random outcomes, and the resource states and measurement patterns can be chosen such that the result is an ensemble of unitary transformations\cite{Someya}.
Here we show that fixed graph states with deterministic measurement patterns can yield ensembles of unitary transformations on an arbitrary input that satisfy the $t$-design condition approximately \emph{and} exactly.
A connection between using classically randomised MB schemes to generate pseudorandomness (in the form of typical entanglement) and approximate unitary $t$-designs was mentioned in the optimization of random circuit constructions\cite{BrownWeinsteinViola}.
The advantage of inherent quantum randomness in MB schemes over random circuits was previously pointed out\cite{PlatoDahlstenPlenio}, also in the context of generating typical entanglement. 
We see here this advantage extends to more general pseudorandomness -- $t$-designs -- in a natural way. 
In addition to the practical benefit of not requiring classical randomness and reconfiguration, the MB approach lends itself to new examples; we report exact MB 3-designs using only five and six qubits, within reach of current experiments, and give evidence of their novel mathematical structure.

\emph{\textbf{Approximate MB unitary designs --}}
Any universal model of computation allows one to implement an arbitrary ensemble of unitaries (or more general processes\cite{NielsenChuang}) as follows.
Consider sampling from the finite ensemble $\{p_i,U_i\}$ ($\sum_i p_i=1$, $p_i \geq 0$, $U_i \in \mathrm{U}(d)$ the set of $d \times d$ unitary matrices), acting on an arbitrary input $|\psi\rangle \in \mathbb{C}^d$.
A bipartite system in the state 
\begin{equation}
\sum_i \sqrt{p_i} |i\rangle \otimes U_{i} |\psi\rangle,
\end{equation}
is created by preparing first  $\sum_i \sqrt{p_i} |i\rangle \otimes \ket{\psi}$, and then applying the controlled operation $\sum_i |i\rangle \langle i| \otimes U_{i}$.
One then performs a projective measurement on the first part in the basis $\{|i\rangle\}$; upon obtaining outcome $j$, unitary $U_j$ is applied to the input, and this occurs with probability $p_j$.
One way to generate pseudorandomness is therefore to take known unitary $t$-design ensembles and apply the above reasoning.
What follows is based on the random circuit construction of Brandao, Harrow and Horodecki\cite{BrandaoHarrowHorodecki} (BHH) and shows that one can implement approximate $t$-designs efficiently using a MB scheme.

We briefly review the BHH construction.
For any matrix $\rho$ on the $t$-fold tensor product of $\mathbb{C}^d$, define its expectation with respect to the Haar measure d$U$ as $\mathbb{E}^t_\mathrm{H}(\rho):=\int \mathrm{d}U \, U^{\otimes t} \rho (U^{\otimes t})^\dag$, where the integral is performed over the entire unitary group U$(d)$.
An ensemble of unitaries $\{p_i,U_i\}$ is an approximate $t$-design if, for all $\rho$, the expectation is `close' to that of the Haar ensemble:
\begin{align}
(1-\epsilon) \mathbb{E}^t_\mathrm{H}(\rho) \leq \sum_i p_i \, U_i^{\otimes t} \rho (U_i^{\otimes t})^\dag \leq (1+\epsilon) \mathbb{E}^t_\mathrm{H}(\rho) , \label{eq:approx}
\end{align}
where for matrices $A \leq B$ if $B-A$ is positive semidefinite, and $\epsilon=0$ for exact designs.

Consider a universal set of two-qubit gates $\mathcal{U} \subset \mathrm{U}(4)$; for technical reasons $\mathcal{U} \ni U$ must contain its inverses $U^\dag$ and the matrix elements of each $U$ must be algebraic.
One constructs a ``parallel'' random circuit on $n$ qubits in steps, at each step performing with probability $1/2$ either the `even' unitary $U_{12} \otimes U_{34} \otimes ... \otimes U_{n-1\,n}$, or the `odd' $U_{23} \otimes U_{45} \otimes ... \otimes U_{n-2\,n-1}$, where each $U_{ij}$ is uniformly randomly sampled from $\mathcal{U}$.
BHH show that for sufficiently many (polynomial in $t$, $n$ and $1/\epsilon$) steps, the ensemble of such circuits is an $\epsilon$-approximate $t$-design.

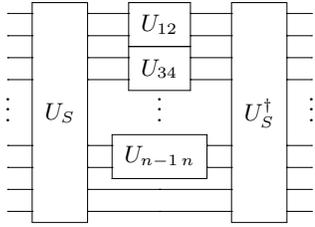
\begin{figure}[h]
\mbox{ 
\Qcircuit @C=1em @R=0em {
 & \multigate{9}{U_S} & \multigate{1}{U_{12}} & \multigate{9}{U_S^\dag} & \qw \\
 & \ghost{U_S} & \ghost{U_{12}} & \ghost{U_S^\dag} & \qw \\
 & \ghost{U_S} & \multigate{1}{U_{34}} & \ghost{U_S^\dag} & \qw \\
 & \ghost{U_S} & \ghost{U_{34}} & \ghost{U_S^\dag} & \qw \\
 \vdots & \pureghost{U_S} & \vdots & \pureghost{U_S^\dag} & \vdots \\
 & \pureghost{U_S} & & \pureghost{U_S^\dag} \\
 & \ghost{U_S} & \multigate{1}{U_{n-1 \, n}} & \ghost{U_S^\dag} & \qw \\
 & \ghost{U_S} & \ghost{U_{n-1 \, n}} & \ghost{U_S^\dag} & \qw \\
 & \ghost{U_S} & \qw & \ghost{U_S^\dag} & \qw \\
 & \ghost{U_S} & \qw & \ghost{U_S^\dag} & \qw
}
}
\caption{ \label{FIG Z(Alt Circuit)} 
One step in the random circuit construction of an approximate $t$-design over $n$ qubits.
The shift gate $U_S$ and its inverse are together either randomly applied or not applied, with the the two-qubit unitaries in between randomly sampled from the universal set $\mathcal{U}$.
Polynomially many iterations of this random circuit will implement an approximate $t$-design\cite{BrandaoHarrowHorodecki}.}
\end{figure}

Starting in an `even' configuration, applying instead an `odd' can be accomplished by a shift operation, defined over the $n$ inputs and two ancilla qubits $n+1$ and $n+2$,
\begin{equation} \label{EQN: Shift}
 U_{S} := S_{n-2\,n+1} S_{n-1\,n+2} \prod_{i=1}^{n-2} S_{i\,i+1},
\end{equation}
where $S_{ij} \in \mathrm{U}(4)$ is the swap operation between qubits $i$ and $j$.
Iterating the circuit described in Fig.~\ref{FIG Z(Alt Circuit)} therefore implements a random parallel circuit.

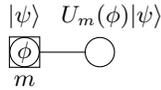
\begin{figure}[!h]
	\centering
	\begin{tikzpicture}[node distance=1cm,main node/.style={circle,draw,inner sep=0.5pt,text centered}]
	\node[main node] (1) {$\phi$};
	\node[draw=black,fit=(1),inner sep=0pt]() {};
	\node[main node] (2) [right of=1] {$\phantom{\phi}$};
	\draw
	(1.north) node[above] {$\ket{\psi}$}
	(1.south) node[below] {$m$}
	(2.north) node[above] {$\quad U_m(\phi)\!\ket{\psi}$};
	\path[every node/.style={font=\sffamily\small}]
	(1) edge node [] {} (2);
	\end{tikzpicture}
	\caption{
The fundamental random unitary transformation induced by measurement on a graph state.
Nodes are qubits initially prepared in the $+1$ eigenstate $|+\rangle $ of the Pauli $X$ operator, and edges indicate entanglement via the controlled-$Z$ ($CZ$) operation.
Angles $\phi$ indicate projective measurement direction in the Pauli $XY$-plane, with the random outcome bit $m$; output nodes are unmeasured and therefore blank.
Here we explicitly include an arbitrary input (square node) state $\ket{\psi}$ and the output; $U_m(\phi)$ is given by Eq.~(\ref{eq:Un}).
} \label{fig:Un}
\end{figure}

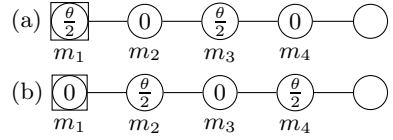
\begin{figure}[h]
	\centering
	\begin{tikzpicture}[node distance=1cm,main node/.style={circle,draw,inner sep=0.5pt,text centered}]
	\node[main node] (1) {$\frac{\theta}{2}$};
	\node[draw=black,fit=(1),inner sep=0pt]() {};
	\node[main node,text width=0.35cm] (2) [right of=1] {$0$};
	\node[main node] (3) [right of=2] {$\frac{\theta}{2}$};
	\node[main node, text width=0.35cm] (4) [right of=3] {$0$};
	\node[main node,text width=0.35cm] (5) [right of=4] {$\phantom{0}$};
	\draw
	(1.west) node[left] {(a)}
	(1.south) node[below] {$m_1$}
	(2.south) node[below] {$m_2$}
	(3.south) node[below] {$m_3$}
	(4.south) node[below] {$m_4$};
	\path[every node/.style={font=\sffamily\small}]
	(1) edge node [] {} (2)
	(2) edge node [] {} (3)
	(3) edge node [] {} (4)
	(4) edge node [] {} (5);
	\end{tikzpicture}
	
	\begin{tikzpicture}[node distance=1cm,main node/.style={circle,draw,inner sep=0.5pt,text centered}]
	\node[main node,text width=0.35cm] (1) {$0$};
	\node[draw=black,fit=(1),inner sep=0pt]() {};
	\node[main node] (2) [right of=1] {$\frac{\theta}{2}$};
	\node[main node,text width=0.35cm] (3) [right of=2] {$0$};
	\node[main node] (4) [right of=3] {$\frac{\theta}{2}$};
	\node[main node,text width=0.35cm] (5) [right of=4] {$\phantom{0}$};
	\draw
	(1.west) node[left] {(b)}
	(1.south) node[below] {$m_1$}
	(2.south) node[below] {$m_2$}
	(3.south) node[below] {$m_3$}
	(4.south) node[below] {$m_4$};
	\path[every node/.style={font=\sffamily\small}]
	(1) edge node [] {} (2)
	(2) edge node [] {} (3)
	(3) edge node [] {} (4)
	(4) edge node [] {} (5);
	\end{tikzpicture}
	\caption{\label{FIG: Z(alpha)}
By measuring the qubits as indicated, (a) implements randomly $Z^{m_1\oplus m_3} X^{m_2\oplus m_4} Z(\theta)^{m_2\oplus 1}$ while (b) implements randomly $Z^{m_3} X^{m_2 \oplus m_4} X(\theta)^{m_3 \oplus 1} Z^{m_1}$, where $\oplus$ denotes bitwise sum (ignoring unimportant global phases).}
\end{figure}

\begin{figure}[h]
\begin{tikzpicture}[node distance=0.75cm,main node/.style={circle,draw,inner sep=0.2pt,text centered}]
  \node[main node,text width=0.35cm] (1) {$0$};
  \node[main node,text width=0.35cm] (2) [right of=1] {$0$};
  \node[main node,text width=0.35cm] (10) [right of=2] {$\phantom{0}$};
   \node[draw=black,fit=(1),inner sep=0pt]() {}; 
  \node[main node,text width=0.35cm] (5) [below of=10] {$\frac{\pi}{4}$};
  \node[main node,text width=0.35cm] (11) [below of=5] {$\phantom{0}$};
  \node[main node,text width=0.35cm] (4) [left of=11] {$0$};
  \node[main node,text width=0.35cm] (3) [left of=4] {$0$};
   \node[draw=black,fit=(3),inner sep=0pt]() {};    
  \node[main node,text width=0.42cm] (6) [right of=5] {$0$}; 
  \node[main node,text width=0.35cm] (7) [right of=6] {$\frac{\pi}{4}$};  
  \node[main node,text width=0.42cm] (8) [right of=7] {$0$}; 
  \node[main node,text width=0.35cm] (9) [right of=8] {$\frac{\pi}{2}$}; 
  \draw
   (1.south) node[below] {$m_1$}
   (2.south) node[below] {$m_2$}
   (3.south) node[below] {$m_3$}
   (4.south) node[below] {$m_4$}
   (5.south) node[left] {$m_5\;$}
   (6.south) node[below] {$m_6$}
   (7.south) node[below] {$m_7$}
   (8.south) node[below] {$m_8$}
   (9.south) node[below] {$m_9$};
  \path[every node/.style={font=\sffamily\small}]
    (1) edge node [] {} (2)
    (2) edge node [] {} (10)
    (3) edge node [] {} (4)
    (4) edge node [] {} (11)
    (5) edge node [] {} (10)  
    (5) edge node [] {} (11)     
    (5) edge node [] {} (6)
    (6) edge node [] {} (7)
    (7) edge node [] {} (8)
    (8) edge node [] {} (9);  
\end{tikzpicture}
\caption{\label{FIG: CNOT}
Graph and measurement pattern implementing the two-qubit gate $U_{ij}=(Z_i Z_j)^{M}( Z(\pi/2)_i Z(\pi/2)_j CZ_{ij})^{m_6 \oplus 1}\times$ $ X^{m_4}_i X^{m_2}_j Z^{m_3}_i Z^{m_1}_j$, where $M$ is a random bit which is a function of measurement results $m_{5,7,8,9}$.
}
\end{figure}
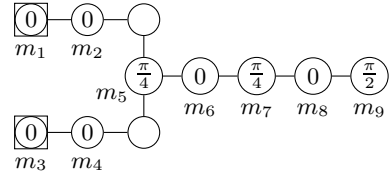

In the remainder of this section we will show how to implement this random parallel circuit with a MB scheme.
The resource state in Fig.~\ref{fig:Un} (written as a graph, see caption) implements the random qubit unitary
\begin{equation} \label{eq:Un}
U_m(\phi) := H Z^m Z(\phi),
\end{equation}
where $m \in \{0,1\}$ is the random measurement outcome, $H$ is the Hadamard matrix, and $Z(\phi):=e^{-i Z \phi/2}$ (similar notation is used for Pauli $X$ and $Y$).
Graphs can be connected (outputs of one identified with the inputs of the next) to perform products of unitaries. 
By connecting several copies of the graph in Fig.~\ref{fig:Un} and choosing measurement angles, Figs. \ref{FIG: Z(alpha)} and \ref{FIG: CNOT} implement certain random one- and two-qubit unitaries, respectively.

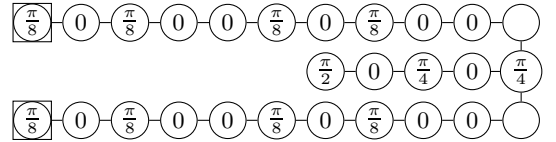
\begin{figure}[h]
\begin{tikzpicture}[node distance=0.65cm,main node/.style={circle,draw,inner sep=0.2pt,text centered}]
  \node[main node,text width=0.35cm] (1) {$\frac{\pi}{8}$};
   \node[draw=black,fit=(1),inner sep=0pt]() {};
  \node[main node,text width=0.42cm] (2) [right of=1] {$0$};
  \node[main node,text width=0.35cm] (3) [right of=2] {$\frac{\pi}{8}$};
  \node[main node,text width=0.42cm] (4) [right of=3] {$0$};
  \node[main node,text width=0.42cm] (5) [right of=4] {$0$};
  \node[main node,text width=0.35cm] (6) [right of=5] {$\frac{\pi}{8}$};
  \node[main node,text width=0.42cm] (7) [right of=6] {$0$};
  \node[main node,text width=0.35cm] (8) [right of=7] {$\frac{\pi}{8}$};
  \node[main node,text width=0.42cm] (9) [right of=8] {$0$};
  \node[main node,text width=0.42cm] (10) [right of=9] {$0$};
  \node[main node,text width=0.42cm] (11) [right of=10] {$\phantom{0}$};
  \node[main node,text width=0.42cm] (12) [below of=11] {$\frac{\pi}{4}$};  
  \node[main node,text width=0.42cm] (13) [below of=12] {$\phantom{0}$};
  \node[main node,text width=0.42cm] (14) [left of=13] {$0$};
  \node[main node,text width=0.42cm] (15) [left of=14] {$0$};
  \node[main node,text width=0.35cm] (16) [left of=15] {$\frac{\pi}{8}$};
  \node[main node,text width=0.42cm] (17) [left of=16] {$0$};
  \node[main node,text width=0.35cm] (18) [left of=17] {$\frac{\pi}{8}$};
  \node[main node,text width=0.42cm] (19) [left of=18] {$0$};
  \node[main node,text width=0.42cm] (20) [left of=19] {$0$};
  \node[main node,text width=0.35cm] (21) [left of=20] {$\frac{\pi}{8}$};
  \node[main node,text width=0.42cm] (22) [left of=21] {$0$};  
  \node[main node,text width=0.35cm] (23) [left of=22] {$\frac{\pi}{8}$};
   \node[draw=black,fit=(23),inner sep=0pt]() {};
  \node[main node,text width=0.42cm] (24) [left of=12] {$0$};   
  \node[main node,text width=0.35cm] (25) [left of=24] {$\frac{\pi}{4}$};
  \node[main node,text width=0.42cm] (26) [left of=25] {$0$};
  \node[main node,text width=0.35cm] (27) [left of=26] {$\frac{\pi}{2}$};
  \draw
    (1) edge node [] {} (2)
    (2) edge node [] {} (3)
    (3) edge node [] {} (4)
    (4) edge node [] {} (5)
    (5) edge node [] {} (6)
    (6) edge node [] {} (7)
    (7) edge node [] {} (8)
    (8) edge node [] {} (9)
    (9) edge node [] {} (10)
    (10) edge node [] {} (11)
    (11) edge node [] {} (12)
    (12) edge node [] {} (13)
    (13) edge node [] {} (14)
    (14) edge node [] {} (15)
    (15) edge node [] {} (16)
    (16) edge node [] {} (17)
    (17) edge node [] {} (18)
    (18) edge node [] {} (19)
    (19) edge node [] {} (20)
    (20) edge node [] {} (21)
    (21) edge node [] {} (22)
    (22) edge node [] {} (23)
    (12) edge node [] {} (24)
    (24) edge node [] {} (25)
    (25) edge node [] {} (26)
    (26) edge node [] {} (27);
\end{tikzpicture}

\caption{\label{FIG: combined}
Measurement gadgets combined in this way sample from a universal set of two-qubit unitaries, given in Eq.(\ref{EQN: two qubit gates}). 
}
\end{figure}

These `gadgets' can be combined to sample from a larger universal set of unitaries; Fig.~\ref{FIG: combined} implements
\begin{align} \label{EQN: two qubit gates}
& U^\mathbf{M}_{ij} = (Z_i Z_j)^{M_1} ( Z(\pi/2)_i Z(\pi/2)_j CZ_{ij})^{M_2} \nonumber \\
& \quad X^{M_3}_i X^{M_4}_j Z^{M_5}_i Z^{M_6}_j Z(\pi/4)^{M_7}_i Z(\pi/4)^{M_8}_j  \\
& \quad X^{M_9}_i X^{M_{10}}_j Z^{M_{11}}_i Z^{M_{12}}_j X(\pi/4)^{M_{13}}_i X(\pi/4)^{M_{14}}_j Z^{M_{15}}_i Z^{M_{16}}_j , \nonumber 
\end{align}
where, here and in the following, $\mathbf{M}$ is a new bit string whose independently random entries are functions of the measurement results $m_k$. 
This set is universal because it contains the universal set $\{ X(\pi/4), Z(\pi/4), CZ \}$; note also that their matrix elements are algebraic.
Furthermore, since $Z X(\pi/4)= X(-\pi/4) Z$, for every $\mathbf{M}$ there exists an $\mathbf{M'}$ such that $U^\mathbf{M'}=(U^\mathbf{M})^{-1}$, thus satisfying the conditions of the BHH construction.

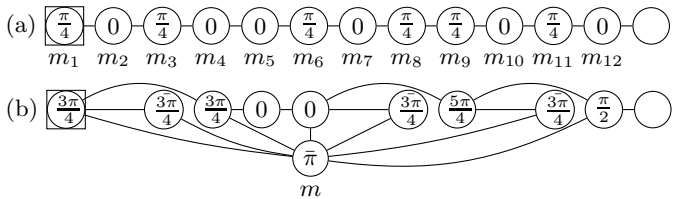
\begin{figure}[h]
\begin{tikzpicture}[node distance=0.65cm,main node/.style={circle,draw,inner sep=0.2pt,text centered}]
  \node[main node,text width=0.35cm] (1) {$\frac{\pi}{4}$};
   \node[draw=black,fit=(1),inner sep=0pt]() {};
  \node[main node,text width=0.42cm] (2) [right of=1] {$0$};
  \node[main node,text width=0.35cm] (3) [right of=2] {$\frac{\pi}{4}$};
  \node[main node,text width=0.42cm] (4) [right of=3] {$0$};
  \node[main node,text width=0.42cm] (5) [right of=4] {$0$};
  \node[main node,text width=0.35cm] (6) [right of=5] {$\frac{\pi}{4}$};
  \node[main node,text width=0.42cm] (7) [right of=6] {$0$};
  \node[main node,text width=0.35cm] (8) [right of=7] {$\frac{\pi}{4}$};
  \node[main node,text width=0.35cm] (9) [right of=8] {$\frac{\pi}{4}$};
  \node[main node,text width=0.42cm] (10) [right of=9] {$0$};
  \node[main node,text width=0.35cm] (11) [right of=10] {$\frac{\pi}{4}$};
  \node[main node,text width=0.42cm] (12) [right of=11] {$0$};
  \node[main node,text width=0.42cm] (13) [right of=12] {$\phantom{0}$};  
  \draw
   (1.west) node[left] {(a)}  
   (1.south) node[below] {$m_1$}
   (2.south) node[below] {$m_2$}
   (3.south) node[below] {$m_3$}
   (4.south) node[below] {$m_4$}
   (5.south) node[below] {$m_5$}
   (6.south) node[below] {$m_6$}
   (7.south) node[below] {$m_7$}
   (8.south) node[below] {$m_8$}
   (9.south) node[below] {$m_9$}
   (10.south) node[below] {$m_{10}$}
   (11.south) node[below] {$m_{11}$}
   (12.south) node[below] {$m_{12}$};   
  \path[every node/.style={font=\sffamily\small}]
    (1) edge node [] {} (2)
    (2) edge node [] {} (3)
    (3) edge node [] {} (4)
    (4) edge node [] {} (5)
    (5) edge node [] {} (6)
    (6) edge node [] {} (7)
    (7) edge node [] {} (8)
    (8) edge node [] {} (9)
    (9) edge node [] {} (10)
    (10) edge node [] {} (11)
    (11) edge node [] {} (12)
    (12) edge node [] {} (13);
\end{tikzpicture}

\begin{tikzpicture}[node distance=0.65cm,main node/.style={circle,draw,inner sep=0.2pt,text centered}]
  \node[main node,text width=0.3cm] (1) {$\frac{3\pi}{4}$};
   \node[draw=black,fit=(1),inner sep=0pt]() {$\phantom{0}$};
  \node[main node,text width=0.29cm] (3) [right of=1,xshift=0.65cm] {$\frac{\bar{3\pi}}{4}$};
  \node[main node,text width=0.29cm] (4) [right of=3] {$\frac{3\pi}{4}$};
  \node[main node,text width=0.42cm] (5) [right of=4] {$0$};
  \node[main node,text width=0.42cm] (6) [right of=5] {$0$};
  \node[main node,text width=0.29cm] (8) [right of=6,xshift=0.65cm] {$\frac{\bar{3\pi}}{4}$};
  \node[main node,text width=0.29cm] (9) [right of=8] {$\frac{5\pi}{4}$};
  \node[main node,text width=0.29cm] (11) [right of=9,xshift=0.65cm] {$\frac{\bar{3\pi}}{4}$};
  \node[main node,text width=0.35cm] (12) [right of=11] {$\frac{\pi}{2}$};
  \node[main node,text width=0.42cm] (13) [right of=12] {$\phantom{0}$};
  \node[main node,text width=0.42cm] (14) [below of=6] {$\bar{\pi}$};    
  \draw
   (1.west) node[left] {(b)}  
   (14.south) node[below] {$m$};   
  \path[every node/.style={font=\sffamily\small}]
    (1) edge node [] {} (3)
    (1) edge [bend left] (4) 
    (4) edge node [] {} (5)
    (5) edge node [] {} (6)
    (6) edge node [] {} (8)
    (6) edge [bend left] (9)
    (9) edge node [] {} (11)
    (9) edge [bend left] (12)
    (12) edge node [] {} (13)
    (1) edge [bend right=5] (14)    
    (3) edge [bend right=10] (14)      
    (4) edge node [] {} (14)    
    (6) edge node [] {} (14)    
    (8) edge node [] {} (14)      
    (11) edge [bend left=5] (14)
    (12) edge [bend left=17] (14);    
\end{tikzpicture}
\caption{\label{FIG: Hadamard}
(a) A naive random $H$, resulting in the unitary given in Eq.(\ref{EQN: Bad HAd}).
(b) A random $H$ where $X$-fusion has imposed common measurement results on the naive case, implementing $X^{M}Z^{M'} H^{m}$ (the random Paulis are harmless; $\bar{\phi}$ indicates measurements in the Pauli $ZY$-plane).}
\end{figure}

We can use these graph gadgets to implement the shift operator of Eq.(\ref{EQN: Shift}).
Each swap can be decomposed into $CZ$ and $H$ gates, which can in turn be decomposed as $H = Z(\pi/2)X(\pi/2)Z(\pi/2)$.
The key observation is that in order to implement a random unitary composed of several gadget unitaries, we must correlate certain random outcomes.
For example, if we naively combined gadgets to try to perform a random Hadamard as in Fig.~\ref{FIG: Hadamard}(a), we would get the random unitary 
\begin{equation} \label{EQN: Bad HAd}
	X^{M_1}Z^{M_2} Z(\pi / 2)^{M_3} X(\pi / 2)^{M_4} Z(\pi / 2)^{M_5}.
\end{equation}
As we will see, the random Paulis on the left can be ignored; however, each of the three rotations are independently randomly applied, failing to implement $H$ most of the time.
We want to set $M_3=M_4=M_5$, and find that $M_3=m_2 \oplus 1$, $M_4 = m_7 \oplus 1$ and $M_5= m_{10} \oplus 1$, so this can be done by projecting qubits $2$, $7$ and $10$ onto the same results in the $X$ basis. 

Projecting a set of vertices onto identical results can be accomplished by a new graph where the set is replaced with a single vertex in a particular way.
In the case of common $Z$ measurements on two qubits this is exactly the ``fusion'' operation of optical MBQC\cite{BrowneRudolph}.
Here we require $X$ ($\phi=0$) measurements to be correlated as these give rise to the crucial dependencies, and we call this graph transformation an $X$-fusion operation; see the 
appendix 
for details.
In the case of the Hadamard example where the set of vertices to be correlated is $\{2,7,10\}$, the resulting graph is given in Fig.~\ref{FIG: Hadamard}(b). 
Note that $X$-fusion introduces local Clifford operations that can change the measurement basis to the $ZY$-plane.

The random unitary resulting from Fig.~\ref{FIG: CNOT} has unwanted $Z(\pi/2)$ rotations correlated to the $CZ$. 
We can now use $X$-fusion to undo this: simply append $Z(\pi/2)$ gadgets (Fig.~\ref{FIG: Z(alpha)}(a)) and impose correlations using appropriate X-fusions, resulting in a new (rather complicated) graph.
We assume this has been done in the following, where we combine these results to construct a graph that implements the random circuit of Fig.~\ref{FIG Z(Alt Circuit)}.

To find the graph for $U_{S}$ we first decompose its circuit description into $Z(\pi/2)$, $X(\pi/2)$ and $CZ$. 
Where $Z(\pi/2)$ and $X(\pi/2)$ appear we use the gadgets of Fig.~\ref{FIG: Z(alpha)}(a) and (b) respectively, and where $CZ$ appears we use the gadget of Fig.~\ref{FIG: CNOT} (adapted as mentioned above).
The same procedure can be used for $U_{S}^\dagger$.
Between each pair of appropriate outputs of $U_S$ and inputs of $U_S^\dag$ we insert the two-qubit gadget of Fig.~\ref{FIG: combined}.
Looking at the induced unitaries corresponding to the gadgets (see figure captions), we see that, because the non-Pauli gates are Clifford, all the random Paulis can be moved to the left; this allows them to be absorbed into the randomly sampled two-qubit unitaries of Eq.(\ref{EQN: two qubit gates}), which remain universal.
It remains to force all of the appropriate random $U_{S}$ and $U_{S}^\dagger$ outcomes to be the same; as in the Hadamard example of Eq.(\ref{EQN: Bad HAd}), ignoring Paulis we have the correct combination of rotations, apart from the fact that they occur independently.
To correlate them we apply $X$-fusions on the appropriate qubits in each of the gadgets that make up the $U_{S}$ and $U_{S}^\dagger$ graphs.
In this way we end up with a large graph, with fixed measurement angles prescribed by the gadgets, that implements the random parallel circuit of Fig.~\ref{FIG Z(Alt Circuit)}. 
Connecting such graphs effects repeated iterations of the random circuit as required.

It only remains to check that the graph does not scale badly in size or preparation time.
The number of qubits used is polynomial in $n$ because the number of gadgets used is linear in the number of BHH's gates, and each gadget has a fixed number of nodes.
The number of edges puts an upper bound on the preparation time.
Each gadget has a fixed number of edges, and linearly many gadgets are used, so we need only be concerned with operations that change edges -- the $X$-fusions.
In the 
appendix 
we show that these can be chosen so that edges do not proliferate.
In this way the number of edges is fixed for each gadget, so the total number of edges is linear in the number of gadgets, and therefore also in $n$. 

This shows that fixed resource states with fixed measurement settings can give rise to pseudorandom ensembles in the form of approximate $t$-designs for all $t$, $n$ and $\epsilon$.
The construction is efficient but requires a large overhead, which we expect can be greatly improved.

\emph{\textbf{Exact linear cluster designs --}}
We will now show that the MB approach can produce exact designs with surprisingly few resources. 
From Eq.(\ref{eq:Un}) it follows that a linear cluster of $L$ qubits
\begin{center}
\begin{tikzpicture}[node distance=1cm,main node/.style={circle,draw,inner sep=0.1pt,text centered}]
  \node[main node] (1) {$\phi_1$};
   \node[draw=black,fit=(1),inner sep=0pt]() {};
  \node[main node] (2) [right of=1] {$\phi_2$};
  \node[] (3) [right of=2] {$\cdots$};
  \node[main node] (4) [right of=3] {$\phi_L$};
  \node[main node] (5) [right of=4] {$\phantom{\phi_1}$};
  \draw
   (1.north) node[above] {$\ket{\psi}$}
   (1.south) node[below] {$m_1$}
   (2.south) node[below] {$m_2$}
   (4.south) node[below] {$m_L$}
   (5.north) node[above] {$\quad U_{\mb{m}}(\mb{\phi})\!\ket{\psi}$};
  \path[every node/.style={font=\sffamily\small}]
    (1) edge node [] {} (2)
    (2) edge node [] {} (3)
    (3) edge node [] {} (4)
    (4) edge node [] {} (5);
\end{tikzpicture}
\end{center}
yields a unitary
\be\label{eq:UL}
U_{\mb{m}}(\mb{\phi}):=U_{m_L}(\phi_L) \cdots U_{m_2}(\phi_2)U_{m_1}(\phi_1),
\ee
where $\mb{\phi}\in[0,\pi]^L$ and $\mb{m}\in\{0,1\}^L$ are ordered lists of angles and outcomes, respectively.
Here node 1 is the input, and node $L+1$ is the output.
We are interested in the ensemble of unitaries $\left\{ p_{\mb{m}}, U_{\mb{m}}(\mb{\phi}) \right\}$ for all outcome strings $\mb{m}$.
The linearity of the cluster ensures that $p_{\mb{m}} = 1/2^L$ will be the same for all $\mb{m}$, and since an ensemble has $2^L$ elements the distribution is uniform.

A test for $t$-designess can be made using the \emph{frame potential}\cite{RenesBlumekohoutScottCaves,GrossAudenaertEisert}, which is a sum of powers of the ensemble elements' Hilbert-Schmidt overlaps. 
In our case of a uniform ensemble on qubits it is given by
\begin{align}\label{eq:tdmin}
F^t_L(\mb{\phi})
&:=\frac{1}{4^L} \sum_{\mb{m},\mb{m}'} \left| \mathrm{Tr}\left[ U_{\mb{m}}(\mb{\phi})^\dag U_{\mb{m}'}(\mb{\phi})\right] \right|^{2t} \geq \frac{(2t)!}{t!(t+1)!} ,
\end{align}
and the bound on the r.h.s. is known to be achieved if and only if the ensemble is a $t$-design.
Equations~(\ref{eq:Un},\ref{eq:UL}) along with the cyclicity of the trace imply that the first and last measurement angles, $\phi_1$ and $\phi_L$, do not affect the frame potential -- note this does not mean the nodes themselves are redundant, since their measurement outcomes help to grow the ensemble.
The frame potential is also symmetric under the transposition $\phi_{l+1} \leftrightarrow \phi_{L-l}$.

A $t$-design is by definition a $(t-1)$-design, and it is not hard to see that a 1-design must span the operator space, thus any design for the unitary group U$(d)$ must contain at least $d^2$ elements.  
Since here $d=2$ and the $L=1$ ensemble has but 2 elements, it cannot be a design.  
For $L=2$ the frame potential is easily computed:  $F^1_2(\mb{\phi}) =1$, which coincides with the minimum in Eq.~(\ref{eq:tdmin}) for all $\mb{\phi}$ and is therefore always a 1-design, (choosing $\mb{\phi}=\{0,0\}$ gives the Pauli ensemble up to phase).  
Any basis is a 1-design, and so we will subsequently concern ourselves with $t\geq2$.

For $L=3$ the frame potential is $F^2_3(\mb{\phi}) = 2(1+\cos^4\phi_2 + \sin^4\phi_2)$ , which has a global minimum of 3 at $\phi_2=\pi/4$; this exceeds the 2-design minimum of 2 from Eq.~(\ref{eq:tdmin}).  
This is not surprising, since there are 8 elements in the ensemble and a lower bound of 10 has been proved\cite{RoyScott}.
For $L=4$, one finds the product $F^2_4(\mb{\phi}) = F^2_3(\phi_2) F^2_3(\phi_3)/4$; each factor can be independently minimised at angle $\pi/4$, yielding $9/4 > 2$.
Thus even though there are more than the minimal number of elements, we have proved that for $L=4$ no choice of angles can give a 2-design, (and hence any $(t\geq 2)$-design). 

For $L=5$ the frame potential can be written
\be
F^2_5(\mb{\phi}) = 4X_2X_4\left(x_3^2 + \left(3(1-X_2^{-1})(1-X_4^{-1})-1\right)x_3+1\right) ,
\ee
where $X_2:=1-\cos^2\phi_2+\cos^4\phi_2$, similarly for $X_4$, and $x_3=\cos^2\phi_3$.
This has a unique minimum of 2 at $X_2=X_4=3/4$ and $x_3=1/3$.
Since this achieves the bound we do indeed have a 2-design, or more precisely a set of (intimately related) 2-designs as there are several choices of equivalent angles, the simplest being $\phi_2=\phi_4=\pi/4$ and $\phi_3=\arccos\sqrt{1/3}$.

One finds that this ensemble is also a 3-design; $F^3_5(\phi_1,\pi/4,\arccos\sqrt{1/3},\pi/4,\phi_5)=5$, again achieving the bound in Eq.~(\ref{eq:tdmin}).
However, the $t=4$ value is $14 \frac{14}{27} > 14$, and so it does not define a 4-design.
We pause here to note that previous design constructions are predominantly related to group actions\cite{GrossAudenaertEisert, RoyScott}, and in particular it is well known that 3-designs are generated by the Clifford group\cite{Dankert, Turner}.
One is led to ask whether or not the 32 unitary matrices (see 
appendix 
) in this $L=5$ qubit 3-design also admit a finite group structure.
Due to the irrationality of $\phi_3$ however, any group containing the ensemble must have infinite order.
Additionally, the number of ensemble elements for any such MB design must be a power of 2, which is not the case for Clifford designs.
Thus it would seem that along with being practically motivated, MB designs are mathematically novel.

The following two facts are not hard to prove: if $\{ p_i , U_i\}$ is a $t$-design, then so is $\{ p_i , V U_i W\}$ for any $V,W \in$ U$(d)$; and the ensemble formed by the (uniform) union of a $t$-design and a $t'$-design is a min$(t,t')$-design.
Together they imply that once a MB $t$-design has been achieved, any choice of subsequent measurement pattern will output at least a $t$-design.
Thus any measurement pattern including the subsequence $\{ 1/2, 1/3, 1/2 \}$ will remain a 3-design, where we have switched to a more natural parameterization $\phi \rightarrow x=\cos^2\phi$.
For $L=6$ calculations can still be carried out analytically, and interestingly a continuous family of 3-designs arises for angles given in the new parameterization by
\begin{align}
\mb{x} = \left\{ x_1, \frac{1}{2}, x_3, \frac{3x_3-2}{3x_3-3}, \frac{1}{2}, x_6 \right\} , \quad x_3 \in \left[ 0,\frac{2}{3}\right].
\end{align}

\begin{figure}[!h]
\centering
\setbox1=\hbox{\includegraphics[width=\columnwidth]{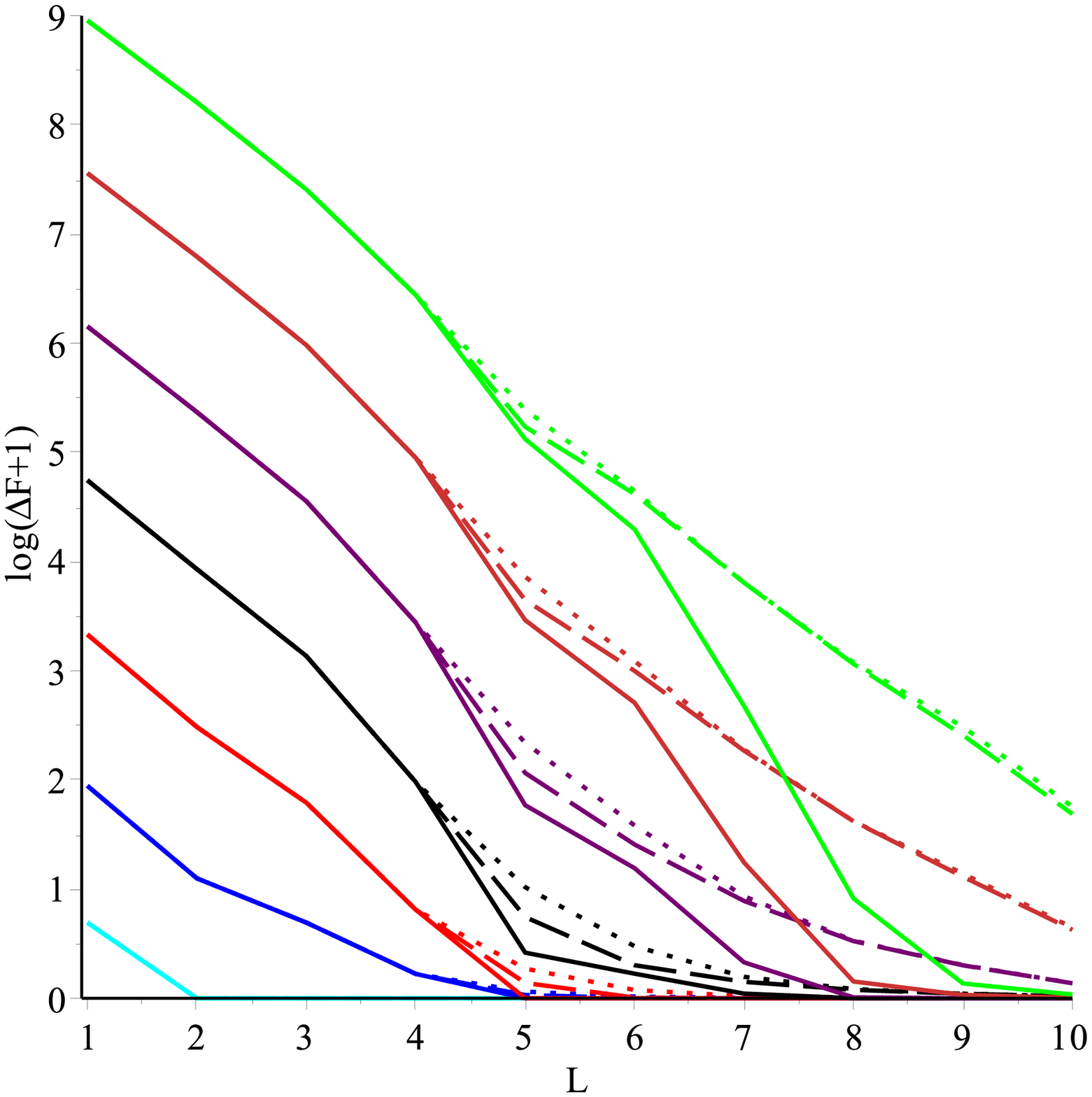}}
\includegraphics[width=\columnwidth]{LinFP.eps}\llap{\raisebox{0.72\columnwidth}{\includegraphics[width=0.6\columnwidth]{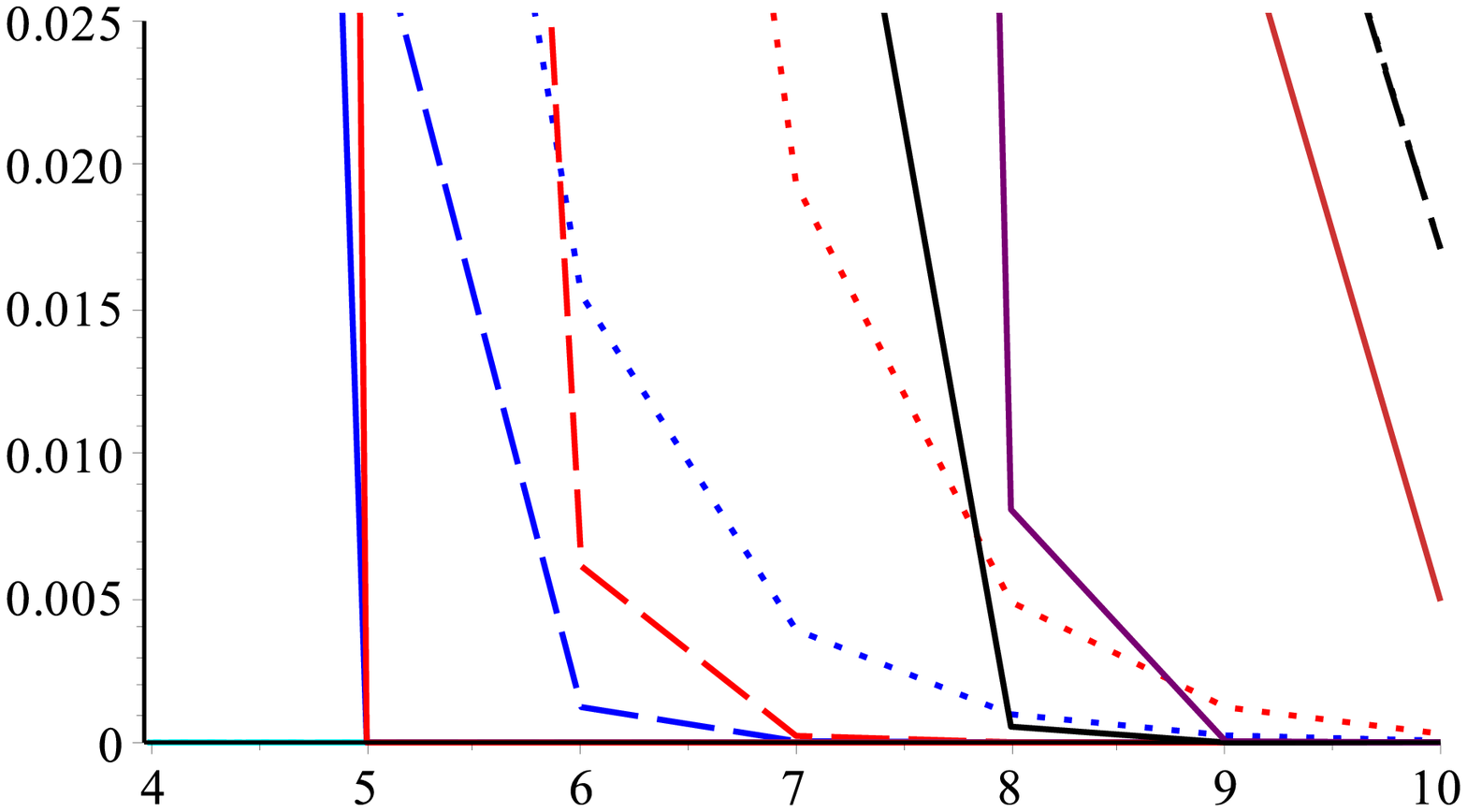}}}
\caption[]{\label{fig:LinFP} 
From bottom to top the $t=1,\cdots,7$ frame potentials (interpolated), given by the difference $\Delta F$ from the exact bound (logarithmic scale) versus linear cluster length $L$.
For each we consider three measurement patterns: dotted lines for those consisting entirely of the angle $\pi/4$; dashed lines for those consisting of a single measurement angle $\phi_\mathrm{min}$ that minimizes the frame potential; and solid lines for a full multi-angle minimization (performed in Matlab).
One sees that the former approach the bound exponentially, albeit with a decreasing rate, as predicted by random quantum circuit results\cite{EmersonLivineLloyd}.
The latter can be seen to drop much more quickly beyond $L=4$.
Other than the trivial $t=1$ case, only the $t=2,3$ curves reach $\Delta F = 0$ (inset), {\it i.e.} the exact design for $L=5$.
Despite the $t=4,5$ curves coming very close to zero, an analytic solution at $L=9$ has not been found\cite{sostools}.
}
\end{figure}

We can carry on the search for higher order designs in longer linear clusters, however the computational demands grow quickly and exact results are elusive.
Figure~\ref{fig:LinFP} shows the difference $\Delta F$ of the first seven frame potentials from the bound for linear clusters up to $L=10$. 
Since the frame potential is the square of a 2-norm\cite{GrossAudenaertEisert}, one finds\cite{Low} that $\sqrt{\Delta F}$ is an upper bound on the diamond norm definition of approximate $t$-designs used in Eq.~(\ref{eq:approx}).
Thus a decreasing frame potential indicates a better approximate $t$-design, and there are several strategies for trying to minimize it.
Figure~\ref{fig:LinFP} shows three such, discussed in the caption.

\emph{\textbf{Multi-qubit cluster designs --}}
The linear cluster results beg the question of the existence of exact MB designs for arbitrary graph states with multi-qubit inputs and outputs, in particular square lattice cluster states of $N$ qubits in $L$ layers.
Note that the tensor product of two $t$-designs is \emph{not} a $t$-design on the tensor product space, which can be seen by recognizing that such a tensor product will never reproduce the entangled correlations of the Haar ensemble.
Thus, two linear cluster 2-designs such as those described above will not give a two-qubit 2-design, as some non-locality will have to be introduced.
A square lattice cluster state can be viewed as doing so by introducing $CZ$ gates between linear clusters.
However, this does not introduce any new free parameters over which we can try to optimize the pseudorandomness of the output ensemble (\textit{e.g.} minimize a frame potential) -- it only introduces non-locality in a very rigid way.
Unfortunately this makes it impossible to find small examples of exact multi-qubit designs.
A numerical exploration of the problem shows that the same general behaviour, (exponential convergence to the Haar value, as in Fig.~\ref{fig:LinFP}), is exhibited by square clusters, but the complexity of the computation prohibits an extensive search.
Clearly the way forward is to identify a (likely group) structure in the ensembles that can be exploited in the multi-qubit case; the exact results above are a major step in this direction, but further investigation is required.

\emph{\textbf{Conclusion --}}
We have shown that there exist MB resources that produce arguably the most randomness possible in the form of approximate and exact $t$-designs.
This arises despite no classical randomness being injected into the system; they are fixed graph states with a deterministic measurement pattern, outputting ensembles that are sampled quantum mechanically.

The role of $t$-designs in quantum estimation\cite{Scott}, in particular randomized benchmarking\cite{benchmarking}, along with cluster states being an important model for error corrected quantum computation in realistic hardware, leads one to anticipate MB designs being useful in the near future.
Related work has recently been done where ancillas in a random circuit model are used to realize exact 2-designs with a quadratic improvement in resources\cite{CleveLeungLiuWang}.
This work demonstrates a new method for finding useful designs, that could make use of powerful MB techniques such as gFlow\cite{BrowneKashefi}.
The broad question raised is, \emph{what resource states provide the most (pseudo)randomness most efficiently?}
In this direction it is intriguing to note that the MB approach can give rise to probability distributions that are impossible to efficiently sample classically\cite{Hoben}, leading one to imagine MB resources that outperform classical randomization in principle as well as in practice.
Several generalizations come to mind, including arbitrary graphs, qu$d$it nodes, non-standard resource preparations ({\it e.g.} $>2$-body entangling gates), and weighted designs. 
We hope this work motivates further research into these and other possibilities.

\emph{\textbf{Acknowledgements --}}
The authors would like to thank D.~Gross, D.~Mahler, T.~Rudolph, A.~Doherty, A.~B.~Sainz, A.~Scott, A.~Roy and S. Bartlett for helpful discussions.
PST acknowledges support from an EPSRC First Grant, US ARO Grant No. W911NF-14-1-0133, and a School of Physics travel grant. 
DM acknowledges support from ANR grant COMB and ville de Paris grant CiQWii.

\bigskip
\appendix
\begin{center}
APPENDIX
\end{center}
\bigskip

\emph{\textbf{Generalised fusion operations--}}
In order to have correlated random unitaries in a measurement based (MB) scheme, we wish certain measurement results to be correlated. 
Say we want to impose the same result on vertices $A=\{a\}$.
To do this, we can think of replacing those vertices with a single vertex, $\alpha$, whose measurement outcome will be this correlated result.
This is done by applying the following projector
\begin{equation}
\sum_m |m\rangle_{\alpha \; A} \langle m,m,\cdots,m| ,
\end{equation}
where $m$ is the measurement result, and $\{|m\rangle_\alpha\}_m$ is the measurement basis on vertex $\alpha$. 
When the measurement basis is Pauli, it turns out this operation can be understood in terms of graph rewrite rules as a generalisation of the ``fusion'' operation\cite{BrowneRudolph}. 
For our gadgets the measurement results that should coincide will always be in the $X$ basis. 

It is useful to review the graphical notation we are using in a more formal way \cite{HeinEisertBriegel, BrowneKashefi}.
Start with a graph $G$ composed of vertices $V$ and edges $E$. 
Each vertex $v \in V$ represents a qubit.
Certain vertices represent the inputs $I \subset V$ (identified by having a box around them) whose qubits are in some state $|\psi\rangle_I$. 
Non-input vertices represent qubits intialised in the state $|+\rangle:=(|0\rangle+|1\rangle)/\sqrt{2}$. 
Edges $E$ represent the application of control-Z gates ($CZ$).
This is sometimes called the open graph state,
\begin{equation} \label{EQN: Open GS}
|G(\psi)\rangle_V = \prod_{(ij) \in E} CZ_{ij} |\psi\rangle_{I}|+\cdots+\rangle_{V/I}.
\end{equation}
To perform a computation, non-output vertices are measured along angles in the Pauli $XY$-plane\cite{BrowneKashefi}; the simplest example is given in Fig.~\ref{fig:Un}.
In order to make a MB computation deterministic, corrections are made to account for random outcomes. 
For example, doing the correction $Z^mH$ on Eq.(\ref{eq:Un}) would implement the deterministic unitary $Z(\phi)$. 
In our situation however we do not want to correct for the measurement results -- indeed they are the source of randomness for our ensembles.

Another way of describing the open graph state of Eq.(\ref{EQN: Open GS}) is via its stabilisers, defined for all noninput vertices $a \notin I$ as 
\begin{equation}
K_a = X_a \bigotimes_{b \in N(a)} Z_b,
\end{equation}
where $N(a)$ indicates the set of neighbours of vertex $a$. Open graph states satisfy the stabiliser equations
\begin{equation}
K_a |G(\psi)\rangle = |G(\psi)\rangle.
\end{equation}
We will also make use of the squareroot stabilisers $\sqrt{K_a}:=X(\pi/2)_a \bigotimes_{b \in N(a)} Z(\pi/2)_b$. 
The following operation takes an open graph state to a new one in which the graph given by the local complementation
\begin{equation}
T_a:=\bigotimes_{b\in N(a)}Z_b \sqrt{K_a}.
\end{equation}
The local complementation of a graph around vertex $a$, denoted $\tau_a$ is given by complementing its neighbourhood, \textit{i.e.} if two neighbours of $a$ are connected in the original graph, they become disconnected, and vice versa. 
This operation is used extensively in quantum information processing using graph states\cite{HeinEisertBriegel},
\begin{equation} \label{EQN: LC}
T_a |G(\psi)\rangle = |\tau_a(G)(\psi)\rangle.
\end{equation}

We begin by considering fusions in the case where all the measurements are in the $Z$ basis, which is a simple extension of the two qubit fusion introduced in \cite{BrowneRudolph}. 
In the case of the other Pauli measurements, local complementation is used to jump between bases as done in \cite{VandenNestetal,HeinEisertBriegel}.
We will only consider fusion projections occurring on non-inputs, and furthermore they will only have non-input neighbours; that is $A \notin I$ and $N(A) \notin I$. 
This is so that the stabiliser relations can be suitably applied, and allows us to treat graphs with no inputs in the proofs for simplicity. 

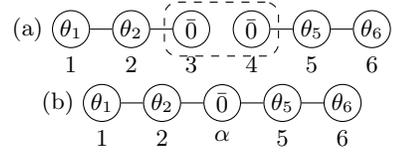
\begin{figure}[h]
\begin{tikzpicture}[node distance=0.8cm,main node/.style={circle,draw,inner sep=1pt,text centered},multi node/.style={rounded rectangle,draw,inner sep=1.8pt,text centered}]
  \node[main node] (1) {$\theta_1$};
  \node[main node] (2) [right of=1] {$\theta_2$};
  \node[main node,text width=0.3cm] (3) [right of=2] {$\bar{0}$};
  \node[main node,text width=0.3cm] (4) [right of=3] {$\bar{0}$};
   \node[draw,dashed,fit=(3)(4),rounded corners=5pt]() {};
  \node[main node] (5) [right of=4] {$\theta_5$};
  \node[main node] (6) [right of=5] {$\theta_6$};
  \draw
   (1.west) node[left] {(a)}  
   (1.south) node[below] {$1$}
   (2.south) node[below] {$2$}
   (3.south) node[below] {$3$}
   (4.south) node[below] {$4$}
   (5.south) node[below] {$5$}
   (6.south) node[below] {$6$}
    (1) edge node [] {} (2)
    (2) edge node [] {} (3)
    (4) edge node [] {} (5)
    (5) edge node [] {} (6);
\end{tikzpicture} \\
\begin{tikzpicture}[node distance=0.8cm,main node/.style={circle,draw,inner sep=1pt,text centered},multi node/.style={rounded rectangle,draw,inner sep=1.8pt,text centered}]
  \node[main node] (1) {$\theta_1$};
  \node[main node] (2) [right of=1] {$\theta_2$};
  \node[main node,text width=0.3cm] (3) [right of=2] {$\bar{0}$};
  \node[main node] (4) [right of=3] {$\theta_5$};
  \node[main node] (5) [right of=4] {$\theta_6$};
  \draw
   (1.west) node[left] {(b)}  
   (1.south) node[below] {$1$}
   (2.south) node[below] {$2$}
   (3.south) node[below] {$\alpha$}
   (4.south) node[below] {$5$}
   (5.south) node[below] {$6$}
    (1) edge node [] {} (2)
    (2) edge node [] {} (3)
    (3) edge node [] {} (4)
    (4) edge node [] {} (5);
\end{tikzpicture}
\caption{\label{FIG: FZ} 
Example of the $Z$ fusion operation, where vertices $3$ and $4$ are to be $Z$-fused. 
A bar indicates measurements in the $ZY$-plane (hence $\bar{0}$ represents a $Z$ basis measurement).
Thus, measuring vertices $A=\{3,4\}$ in the $Z$ basis in graph (a) and imposing the same results is equivalent to measuring vertex $\alpha$ in in the $Z$ basis in graph (b).}
\end{figure}

We start with a simple expansion for any graph state,
\begin{equation} \label{EQN: GZexpn}
|G\rangle_V = \sum_{\mathbf{m}} |\mathbf{m}\rangle_A \prod_{a \in A} Z_{N(a)}^{m_a} |g\rangle_{V/A},
\end{equation}
where here the $|\mathbf{m}\rangle_A$ is a product state in the computational basis for bit string $\mathbf{m}$ of length $|A|$, $g$ is the subgraph given by removing all the vertices in $A$ and attached edges, $m_a$ is the measurement outcome for node $a$, and $Z_{N(a)}$ is shorthand for applying $Z$ on the vertices $N(a)$.
We define the $Z$ basis fusion on vertices $A$ as
\begin{equation}
F_Z^{A} := |0\rangle_{\alpha \; A} \langle 00\cdots0| + |1\rangle_{\alpha \; A} \langle 11\cdots1|.
\end{equation}
From the expansion Eqn.~(\ref{EQN: GZexpn}), it is clear that this has the effect
\begin{equation}
F_Z^{A} |G\rangle = \sum_{m} |m\rangle_\alpha  Z_{\Delta_{a \in A}N(a)}^m |g\rangle_{V/A},
\end{equation}
where $\Delta_{a \in A}N(a)$ denotes the $n$-fold symmetric difference over the sets $N(a)$, $a \in A$ \cite{Someya}. 
Graphically this is just the set of vertices which are connected an odd number of times to $A$. 
The resulting state is also a graph state, found in two steps. 
First add a new vertex $\alpha$ and connect it to the odd neighbourhood of $A$ (again, this is given by the symmetric difference of all the neighbours of $a \in A$). 
Second remove vertices $A$ and all their edges.
We denote the new graph as $F_Z^A (G)$.
In the case that $A = \{a_1, a_2\}$ is composed of two vertices, the new graph is found by simply replacing the two vertices by a new vertex $\alpha$ which is connected to the neighbours of $a_1$ and $a_2$, minus the neighbours common to both. 
See for example Fig.~\ref{FIG: FZ}. 
This is exactly the fusion operation used in \cite{BrowneRudolph}.

To see how the remaining Pauli basis fusions work, we use the fact that the bases can be related to each other by Clifford operations, which in turn can be mapped to graphical operations through local complementation \cite{HeinEisertBriegel}: 
\begin{equation}
\begin{split}
|+\rangle &= Y(\pi/2) |0\rangle  \\ 
& = e^{-i\pi/4}Z(-\pi/2)X(-\pi/2) |0\rangle\\
|-\rangle &= Y(\pi/2) |1\rangle  \\
& = - i e^{-i\pi/4}Z(-\pi/2)X(-\pi/2) |1\rangle\\
|+i\rangle &= X(-\pi/2) |0\rangle  \\
|-i\rangle &= -i X(-\pi/2) |1\rangle,
\end{split}
\end{equation}
where $|\pm (i) \rangle :=  (|0\rangle \pm (i) |1\rangle)/\sqrt{2}$. 
So for the $X$ fusion projection $F_X^{A} := |+\rangle_{\alpha \; A} \langle ++\cdots+| + |-\rangle_{\alpha \; A} \langle --\cdots-|$, we have
\begin{equation} \label{EQN: FX}
F_X^{A} =  Y(\pi/2)_\alpha Z(-\pi |A|/2)F_Z^{A} \bigotimes_{a \in A} X(\pi/2)_a Z(\pi/2)_a.
\end{equation}

To relate this to the local complementation of Eq.~(\ref{EQN: LC}), we note that for two non-input neighbours $a,b \notin I$, 
\begin{widetext}
\begin{align}
X_a(\pi/2)Z_a(\pi/2) =  \left( Y_a\otimes X_b(-\pi/2)Z_b(\pi/2)
\bigotimes_{c\in N(a)\Delta N(b)}Z_c(\pi/2) \bigotimes_{d \in N(a) \cap N(b)} Z_d \right) T_a T_b,
\end{align}
\end{widetext}
where $A \Delta B$ indicates the symmetric difference between sets $A$ and $B$. 
Then we observe than $F_Z^A \bigotimes_{a\in A} Y_a = i^{|A|}  X_\alpha Z^{|A|}_\alpha F_Z^A$.
Next, for each vertex $a \in A$ we choose a neighbour $b^a \in N(a)$. 
If this can be done such that $b^a \notin A$ and $N(b^a)\cap A = a$ (as is the case for all our gadgets -- in other cases similar rules can be found using the same reasoning), $F_X^{A}$ can be given the simple form
\begin{widetext}
\begin{equation}\label{EQN: Fx LUs}
F_X^A = Y(\pi/2)_\alpha Z(\pi |A| /2)_\alpha X_\alpha  
\prod_{a \in A}\left( X_{b^a}(-\pi/2)Z_{b^a}(\pi/2)
\bigotimes_{c\in N(a)\Delta N(b^a)}Z_c(\pi/2) \bigotimes_{d \in N(a) \cap N(b^a)} Z_d \right)
F_Z^{A} \prod_{a \in A} T_a T_{b^a}.
\end{equation}
\end{widetext}

Similarly, for $Y$-fusion  
\begin{equation}
\begin{split}
F_Y^{A} &=  X(\pi/2)_\alpha Z(-\pi |A| /2)_\alpha  F_Z^{A} \bigotimes_{a \in A} X(\pi/2)_a\\
&=  T_\alpha Z(-\pi |A| /2)_\alpha  F_Z^{A} \prod_{a \in A} T_a.
\end{split}
\end{equation}

Remembering that $T_a$ has the effect of implementing a local complementation, the above can be used to find the graphical rules for the fusion projections. 
Up to local unitaries, the graphs after the $X$ and $Y$ fusions are
\begin{eqnarray}
F_X^A(G) &=& F_Z^A(\circ_{a\in A}(\tau_a\circ \tau_{b^a \in N(a)}(G))) \label{EQN: graph FX} , \\
F_Y^A(G) &=& F_Z^A(\circ_{a\in A}(\tau_a (G))) ,
\end{eqnarray}
where $\circ_{a\in A}$ indicates the composition of operations over $A$.

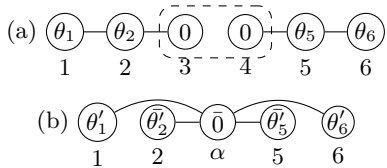
\begin{figure}[h]
\begin{tikzpicture}[node distance=0.8cm,main node/.style={circle,draw,inner sep=1pt,text centered},multi node/.style={rounded rectangle,draw,inner sep=1.8pt,text centered}]
  \node[main node] (1) {$\theta_1$};
  \node[main node] (2) [right of=1] {$\theta_2$};
  \node[main node,text width=0.3cm] (3) [right of=2] {$0$};
  \node[main node,text width=0.3cm] (4) [right of=3] {$0$};
   \node[draw,dashed,fit=(3)(4),rounded corners=5pt]() {};
  \node[main node] (5) [right of=4] {$\theta_5$};
  \node[main node] (6) [right of=5] {$\theta_6$};
  \draw
   (1.west) node[left] {(a)}  
   (1.south) node[below] {$1$}
   (2.south) node[below] {$2$}
   (3.south) node[below] {$3$}
   (4.south) node[below] {$4$}
   (5.south) node[below] {$5$}
   (6.south) node[below] {$6$}
    (1) edge node [] {} (2)
    (2) edge node [] {} (3)
    (4) edge node [] {} (5)
    (5) edge node [] {} (6);
\end{tikzpicture} \\
\begin{tikzpicture}[node distance=0.8cm,main node/.style={circle,draw,inner sep=0pt,text centered},multi node/.style={rounded rectangle,draw,inner sep=1.8pt,text centered}]
  \node[main node,text width=0.41cm] (1) {$\theta_1'$};
  \node[main node] (2) [right of=1] {$\bar{\theta_2'}$};
  \node[main node,text width=0.4cm] (3) [right of=2] {$\bar{0}$};
  \node[main node] (4) [right of=3] {$\bar{\theta_5'}$};
  \node[main node] (5) [right of=4] {$\theta_6'$};
  \draw
   (1.west) node[left] {(b)}  
   (1.south) node[below] {$1$}
   (1.north) node[above] {}
   (2.south) node[below] {$2$}
   (2.north) node[above] {}
   (3.south) node[below] {$\alpha$}
   (3.north) node[above] {}
   (4.south) node[below] {$5$}
   (4.north) node[above] {}
   (5.south) node[below] {$6$}
   (5.north) node[above] {}
    (1) edge [bend left] (3)
    (2) edge node [] {} (3)
    (3) edge node [] {} (4)
    (3) edge [bend left] (5);
\end{tikzpicture}
\caption{\label{FIG: FX} 
Example of an $X$ fusion operation. Here, $\theta_v' := \theta_v +\frac{\pi}{2}$, and again a bar indicates measurement in the $ZY$-plane.
Vertices $3$ and $4$ in (a) are fused according to Eq.~(\ref{EQN: graph FX}), resulting in (b) with local unitaries indicated above. 
Thus, measuring vertices $A=\{3,4\}$ in the $X$ basis in graph (a) and imposing the same results is equivalent to measuring vertex $\alpha$ in the $Z$ basis in (b).}
\end{figure}

Thus $X$-fusion has the effect of changing the graph and applying local unitaries. 
An example of an $X$-fusion can be found in Fig.~\ref{FIG: FX}. 
The graph changes according to the the local complementation rules of Eq.~(\ref{EQN: graph FX}). 
The local unitaries are given by Eq.~(\ref{EQN: Fx LUs}): 
\begin{equation}
\begin{split}
Z_1(\pi/2)\\
X_2(-\pi/2)Z_2(\pi/2)\\
X_5(-\pi/2)Z_5(\pi/2)\\
Z_6(\pi/2)\\
Y_\alpha(\pi/2)Z_\alpha X_\alpha .
\end{split}
\end{equation}
In the figure these are represented by changes to the measurement angles. 
Note that the $X_2(-\pi/2)Z_2(\pi/2)$ rotate the axis of measurement from the $XY$-plane to the $ZY$-plane.

\emph{\textbf{Scaling of approximate MB $t$-designs --}}
First we note that the number of nodes in the MB construction is linear in the number of qubits in the BHH circuit construction.
This can be seen by noting that the shift operation of Eq.~(\ref{EQN: Shift}) involves $n+2$ swap operations, each of which can be decomposed into 3 $CZ$ and 6 $H$ gates, and the latter further decomposes into 3 rotations, giving a total of 21 gadgets per $U_S$ (and similarly $U_S^\dag$).
Furthermore, each $U_{ij}$ in Fig.~\ref{FIG Z(Alt Circuit)} corresponds to a single (appropriately $X$-fused) gadget.
Since the number of nodes in any gadget is fixed, the total number of nodes in the MB graph is linear in $n$.

We also want to show that the $X$-fusion operations do not introduce inefficiencies in the preparation of the graph states. 
Since the number of qubits goes down in the fusion operation, we are only concerned with making sure the number of edges does not grow too quickly. 
The only way that the number of edges could grow is if the local complementations of one gadget affect other gadget graphs. 
It turns out that for the vertices $a\in A$ in the fused set we can choose $b \in N(a)$ in such a way to avoid this.
Concretely, if we consider the local complementations that occur for our gadgets, we can do the following.
For Fig.~\ref{FIG: Z(alpha)}(a) we have $a = 2$ and use $b=3$, for Fig.~\ref{FIG: Z(alpha)}(b) we have $a = 3$ and use $b=2$, and for Fig.~\ref{FIG: CNOT} we have $a=6$ and use $b= 7$.
Since in each case neither vertex is an output, the complementations do not `reach' beyond the gadget. 

BHH show that the random circuit of Fig.~\ref{FIG Z(Alt Circuit)} applied polynomially many times gives an approximate $t$-design. 
More precisely, they show there exists a constant $C(\mathcal{U})$, which depends on the universal set of gates $\mathcal{U}$ used, such that repeating the circuit in Fig.~\ref{FIG Z(Alt Circuit)} $C(\mathcal{U})\lceil log_2(4t) \rceil^2 t^5 t^{3.1}(nt + log(1/\epsilon))$ times forms an $\epsilon$-approximate $t$-design.
Applying this to our measurement based graph state construction where we use a particular universal set, and recalling that the graph state size scales linearly with $n$, we arrive at the following assertion:

The graph construction presented, with the fixed measurement settings detailed, samples from an $\epsilon$-approximate $t$-design. 
Furthermore, there exists a constant $C$ such that the size of the graph is $C \lceil log_2(4t) \rceil^2 t^5 t^{3.1}(nt + log(1/\epsilon))$.

\emph{\textbf{Minimal exact linear cluster design --}}
The 32 elements of the (essentially) unique $L=5$ MB 3-design: 
\begin{widetext}
\begin{align}
  & 
\left[ \begin {array}{cc} 1&0\\ \noalign{\medskip}0&1\end {array} \right] , 
\left[ \begin {array}{cc} 0&1\\ \noalign{\medskip}1&0\end {array} \right] , 
\left[ \begin {array}{cc} 0&i\\ \noalign{\medskip}-i&0\end {array} \right] , 
\left[ \begin {array}{cc} -i&0\\ \noalign{\medskip}0&i\end {array} \right] , \nonumber \\
 \frac{1}{\sqrt{2}} \Bigg\{ &
\left[ \begin {array}{cc} 1&1\\ \noalign{\medskip}1&-1\end {array} \right] , 
\left[ \begin {array}{cc} 1&-1\\ \noalign{\medskip}1&1 \end {array} \right] , 
\left[ \begin {array}{cc} i&-i\\ \noalign{\medskip}-i&-i\end {array} \right] , 
\left[ \begin {array}{cc} -i&-i\\ \noalign{\medskip}i&-i\end {array} \right] \Bigg\} ,\nonumber \\
 \frac{1}{\sqrt{3}} \Bigg\{ &
\left[ \begin {array}{cc} -1&1+i\\ \noalign{\medskip}1-i&1\end {array} \right] , 
\left[ \begin {array}{cc} 1-i&1\\ \noalign{\medskip}-1&1+i\end {array} \right] , 
\left[ \begin {array}{cc} 1+i&i\\ \noalign{\medskip}i&1-i\end {array} \right] , 
\left[ \begin {array}{cc} i&1-i\\ \noalign{\medskip}1+i&i\end {array} \right] \Bigg\} , \nonumber \\
 \frac{1}{\sqrt{6}} \Bigg\{ &
\left[ \begin {array}{cc} -i&2+i\\ \noalign{\medskip}-2+i&i\end {array} \right] , 
\left[ \begin {array}{cc} -2+i&i\\ \noalign{\medskip}-i&2+i\end {array} \right] , 
\left[ \begin {array}{cc} -1-2\,i&-1\\ \noalign{\medskip}-1&1-2\,i\end {array} \right] , 
\left[ \begin {array}{cc} -1&1-2\,i\\ \noalign{\medskip}-1-2\,i&-1\end {array} \right] , \nonumber \\
 &
\left[ \begin {array}{cc} \sqrt{3}-i &-1+i\\ \noalign{\medskip}1+i& \sqrt{3}+i \end {array} \right] , 
\left[ \begin {array}{cc} 1+i& \sqrt{3}+i\\ \noalign{\medskip} \sqrt{3}-i &-1+i\end {array} \right] ,
\left[ \begin {array}{cc} -1+i&-1+i\sqrt {3}\\ \noalign{\medskip}-1-i\sqrt {3}&1+i\end {array} \right] , 
\left[ \begin {array}{cc} -1-i\sqrt {3}&1+i\\ \noalign{\medskip}-1+i&-1+i\sqrt {3}\end {array} \right] , \nonumber \\
 &
\left[ \begin {array}{cc} 1-i\sqrt {3} &-1-i\\ \noalign{\medskip}-1+i&-1-i\sqrt {3}\end {array} \right] , 
\left[ \begin {array}{cc} -1+i&-1-i\sqrt {3}\\ \noalign{\medskip}1-i\sqrt {3}&-1-i\end {array} \right] , 
\left[ \begin {array}{cc} -1-i&\sqrt {3}-i\\ \noalign{\medskip}-\sqrt {3}-i&-1+i\end {array} \right] , 
\left[ \begin {array}{cc} -\sqrt {3}-i&-1+i\\ \noalign{\medskip}-1-i&\sqrt {3}-i\end {array} \right] \Bigg\} , \nonumber \\
 \frac{1}{\sqrt{12}} \Bigg\{ &
\left[ \begin {array}{cc} \omega_+&\omega_-+2i\\ \noalign{\medskip}\omega_--2i&-\omega_+\end {array} \right] , 
\left[ \begin {array}{cc} \omega_--2i&-\omega_+\\ \noalign{\medskip}\omega_+&\omega_-+2i\end {array} \right] , 
\left[ \begin {array}{cc} 2+i\omega_-&-i \omega_+ \\ \noalign{\medskip}-i\omega_+ &2-i\omega_-\end {array} \right] , 
\left[ \begin {array}{cc} -i\omega_+ &2-i\omega_-\\ \noalign{\medskip}2+i\omega_-&-i\omega_+ \end {array} \right] ,\nonumber \\
 &
\left[ \begin {array}{cc} -i\omega_- &-2-i\omega_+\\ \noalign{\medskip}2-i\omega_+&i\omega_- \end {array} \right] , 
\left[ \begin {array}{cc} 2-i\omega_+&i\omega_- \\ \noalign{\medskip}-i \omega_- &-2-i\omega_+ \end {array} \right] , 
\left[ \begin {array}{cc} \omega_++2i&-\omega_-\\ \noalign{\medskip}-\omega_-&-\omega_++2i\end {array} \right] , 
\left[ \begin {array}{cc} -\omega_-&-\omega_++2i\\ \noalign{\medskip}\omega_++2i&-\omega_-\end {array} \right] \Bigg\} ,
\end{align}
\end{widetext}
where each row is an orthonormal Hilbert-Schmidt basis, and we've defined $\omega_\pm = \sqrt{3}\pm 1$.
Ensembles resulting from the removal of any basis fail to be a 2-design.

\emph{\textbf{Partial recursion for the frame potential --}}
The ensemble elements' Hilbert-Schmidt overlaps
\begin{equation}
\langle\langle \mb{m} \vert \mb{m}' \rangle\rangle_{\mb{\phi}} := \mathrm{Tr}\left[ U_{\mb{m}}(\mb{\phi})^\dag U_{\mb{m}'}(\mb{\phi})\right] ,
\end{equation}
define a $2^L \times 2^L$ Gram matrix.
Substituting Eq.(\ref{eq:Un}) into Eq.(\ref{eq:UL}) one finds that due to reductions to previous cases, the important upper triangular Gram elements are of the form $\llg 0 \mb{m} 0 | 1 \mb{m}' 1 \rrg$, where now $\mb{m},\mb{m}'$ are bit strings of length $L-2$ and the angular dependence is implicit.
Let $\mb{\phi}^l_k = \phi_k,\phi_{k+1},\cdots,\phi_{k+l-1}$ and define
\be
f^t(\mb{\phi}^{L-2}_2):= 2 \sum_{\mb{m}} \Big( \vert\llg 0 \mb{m} 0 | 1 \mb{m} 1 \rrg\vert^{2t}
 + 2 \sum_{\mb{m}'>\mb{m}} \vert\llg 0 \mb{m} 0 | 1 \mb{m}' 1 \rrg\vert^{2t} \Big)
\ee
(recall that the angles $\phi_1$ and $\phi_L$ are irrelevant for the frame potential).
Combined with the diagonal ($f^t=0$) cases $F^t_1=4^t/2$ and $F^t_2=4^t/4$, one arrives at a partially recursive formula for the frame potential:
\begin{align}
F^t_{L+1}(\mb{\phi}^{L-1}_2)
 &= 2^{-1}\left[ F^t_L(\mb{\phi}^{L-2}_2) + F^t_L(\mb{\phi}^{L-2}_3) \right] \nonumber\\
 & \quad - 2^{-2} F^t_{L-1}(\mb{\phi}^{L-3}_3) + 2^{-2L-1} f^t(\mb{\phi}^{L-1}_2).
\end{align}
This can be used to reduce the complexity of frame potential calculations, and can give insight into their minimisation.


\begin{thebibliography}{99}   

\bibitem{benchmarking}
J. Emerson, Y. S. Weinstein, M. Saraceno, S. Lloyd and D. G. Cory. Science, 302(5653):2098, (2003);
J. M. Epstein, A. W. Cross, E. Magesan, and J. M. Gambetta, Phys. Rev. A 89, 062321 (2014).

\bibitem{Hayden..Winter}
P. Hayden, D. Leung, P. W. Shor and A. Winter, Comm. Math. Phys. 250, 371, (2004).

\bibitem{Muller..Wiebe}
M. P. Muller, E. Adlam, L. Masanes and N. Wiebe, Comm. Math. Phys. 340, 499 (2015).

\bibitem{Matthews..Turner}
J. C. F. Matthews, R. Whittaker, J. L. O'Brien and P. S. Turner, Phys. Rev. A 91, 020301(R) (2015).

\bibitem{HaydenPreskill}
P. Hayden and J. Preskill, J. High E. Phys. 2007(09), 120 (2007).

\bibitem{DelsarteGoethalsSeidel}
P. Delsarte, J. Goethals, and J. Seidel, Geom. Dedicata, vol. 6, pp. 363 (1977).

\bibitem{RenesBlumekohoutScottCaves}
J.M. Renes, R. Blume-Kohout, A. J. Scott and C. M. Caves, J. Math. Phys. 45, 2171 (2004).

\bibitem{Dankert}
C. Dankert, R. Cleve, J. Emerson and E. Livine, Phys. Rev. A 80, 012304 (2012).

\bibitem{Appleby-Durt}
D. M. Appleby, C. A. Fuchs and H. Zhu, Q. Info. \& Comp. 15, 61 (2015);
T. Durt, B-G. Englert, I. Bengtsson and K. Zyczkowski, Int. J. Quant. Info. 8, 535 (2010).

\bibitem{RaussendorffBriegel}
R. Raussendorff and H. J. Briegel, Phys. Rev. Lett. 86, 5188 (2001).

\bibitem{Someya}
M. Mhalla, M. Murao, S. Perdrix, M. Someya and P. S. Turner, \texttt{arXiv:1006.2616}.

\bibitem{BrownWeinsteinViola}
W. G. Brown, Y. S. Weinstein and L. Viola, Phys. Rev. A 77, 040303(R) (2008).

\bibitem{PlatoDahlstenPlenio}
A.D. Plato, O.C. Dahlsten and M.B. Plenio, Phys. Rev. A 78, 042332 (2008).

\bibitem{NielsenChuang}
M. Nielsen and I. Chuang, ``Quantum Computation and Quantum Information,'' Cambridge 2000.

\bibitem{BrandaoHarrowHorodecki}
F.G.S.L. Brandao, A.W. Harrow and M. Horodecki, \texttt{arXiv:1208.0692}.

\bibitem{BrowneRudolph}
D. Browne and T. Rudolph, Phys. Rev. Lett. 95, 010501 (2005).


\bibitem{HeinEisertBriegel}
M. Hein, J. Eisert and H. J. Briegel, Phys. Rev. A 69, 062311 (2004).

\bibitem{GrossAudenaertEisert} 
D. Gross, C. Audenaert and J. Eisert, J. Math. Phys. 48, 052104 (2007).

\bibitem{RoyScott} 
A. Roy and A. J. Scott, Des. Codes Cryptogr. 53, 13-31 (2009).

\bibitem{Turner}
P. S. Turner, Proceedings, Nankai Series in Pure, App. Math. and Theo. Phys. 11, World Scientific, 2013;
R. Kueng and D. Gross, \texttt{arXiv:1510.02767}.

\bibitem{Low}
R. A. Low, PhD thesis, University of Bristol, (2010) \texttt{arXiv:1006.5227}.

\bibitem{EmersonLivineLloyd}
J. Emerson, E. Livine and S. Lloyd, Phys. Rev. A 72, 060302(R) (2005)

\bibitem{sostools}
Conversely, proving the \emph{non}existence of exact designs should be possible using sum-of-squares techniques for bounding the global minima of polynomials, because these have semi-definite programming certificates; however, the problem seems to be numerically unstable and we were unable to coax convincing bounds on the frame potential from SOStools (\texttt{www.cds.caltech.edu/sostools/}).

\bibitem{Scott}
A. J. Scott, J. Phys. A:  Math. Theor. 41, 055308 (2008).

\bibitem{CleveLeungLiuWang}
R. Cleve, D. Leung, L. Liu and C. Wang, \texttt{arXiv:1501.04592}.

\bibitem{BrowneKashefi}
D.~E.~Browne, E.~Kashefi, M.~Mhalla and S.~Perdrix, New J. Phys. 9 250 (2007).

\bibitem{Hoben}
M.~Hoban, J.~Wallman, H.~Anwar, N.~Usher, R.~Raussendorf and D.~Browne, Phys. Rev. Lett. 112, 140505 (2014).

\bibitem{VandenNestetal}
M. Van den Nest, J. Dehaene, and B. De Moor, Phys. Rev. A 69, 022316 (2004); Phys. Rev. A 70, 034302 (2004).

\end{thebibliography}
\end{document}